# Cross-Modal Semantic Successive Refinement for Generative Image Communication over Underwater Acoustic Channels

Xu Pei, Bin Wang, *Member, IEEE*, Guangxu Zhu, *Member, IEEE*, Yi Zhang, *Member, IEEE* and Haiyan Fan

***Abstract* — Underwater acoustic communication faces challenges such as limited available bandwidth and complex channel environments, which impose demanding requirements on efficient image transmission and reliable reconstruction. To address these challenges, this paper proposes a generative image communication framework based on cross-modal semantic successive refinement. At the transmitter, text descriptions, compact bottleneck features, and visual side information compressed by residual vector quantization are transmitted. At the receiver, text and bottleneck features are first fused to predict a base estimate of the intermediate visual representation. The received RVQ indices are then used to recover supplementary visual information and perform residual correction, followed by image reconstruction using a one-step Pixel MeanFlow model. During training, a CLIP-based text–image semantic consistency constraint is introduced to preserve the semantic correspondence between generated results and the original text descriptions. During transmission, protection resources are allocated according to the importance of different information branches and RVQ quantization stages. Simulations using COCO 2017 images and measured underwater acoustic channel records from Watermark show that text and bottleneck features provide complementary semantic and structural conditions, RVQ visual side information further improves reconstruction quality, and layered unequal error protection helps improve recovery at low SNRs. Code-rate allocation experiments reveal a tradeoff between the reliability of fundamental conditions and visual-detail reconstruction under a limited visual-feature coding budget, providing guidance for allocating protection resources in underwater acoustic image transmission.**

## I. Introduction

Semantic communication (SC) has attracted increasing attention because it focuses on conveying task-relevant meaning rather than ensuring bit-level recovery. Shannon's information theory established the foundations of source compression and reliable channel transmission [1]. Weaver further distinguished the technical, semantic, and effectiveness levels of communication [2]. Early studies explored the quantification of semantic information [3], followed by Bao et al.'s work on semantic information measurement and coding [4]. More recent research has advanced SC in semantic representation, communication architectures, and task-oriented transmission, with increasing attention to contextual information and data significance [5], [6], [7], [8], [9], [10], [11], [12]. In essence, SC aims to identify and transmit information relevant to the receiver's task, enabling the intended content to be reconstructed or interpreted from the received semantic representation and available context.

Underwater visual information is important for applications such as oceanographic observation, marine biology, environmental monitoring, and underwater exploration [13]. Because radio-frequency signals attenuate rapidly in water, acoustic waves are commonly used for underwater wireless communication. However, underwater acoustic channels are characterized by limited bandwidth, long propagation delays, multipath propagation, and Doppler effects, which make reliable image transmission challenging [13]. These constraints motivate the development of communication methods that reduce transmission requirements while preserving important visual information. Semantic communication provides a promising approach by prioritizing semantically relevant content.

Despite the potential of semantic communication for underwater image transmission, several fundamental challenges remain. First, images contain rich and diverse visual information, making it difficult to obtain compact representations while preserving the important information required for accurate reconstruction. Second, the constrained and time-varying nature of underwater acoustic channels makes the reliable transmission of visual information challenging, as channel impairments may lead to information loss and degradation. Third, reconstructing high-quality images from the information available at the receiver remains difficult, particularly when the transmitted representations are incomplete or distorted. Addressing these challenges requires effective approaches to visual representation, reliable information transmission, and image reconstruction in underwater acoustic environments.

To address these challenges, researchers have developed learning-based methods for image representation and transmission. Bourtsoulatze et al. proposed DeepJSCC, which uses a CNN encoder and decoder to jointly encode images and transmit them over a channel [14]. Later studies extended this framework by adding receiver feedback [15], combining it with OFDM [16], introducing attention modules [17], and adapting the transmission bandwidth [18]. Dai et al. proposed NTSCC, which uses learned nonlinear transforms and an entropy model to represent image information for transmission [19]. In parallel, different neural architectures have been explored to improve semantic feature extraction. CNNs are effective in capturing local visual structures, while Transformer-based models can capture long-range dependencies. For image communication, SwinJSCC employs a hierarchical Swin Transformer architecture to extract visual features for joint source-channel coding [20].To support digital transmission, some studies have explored discrete representations. DeepJSCC-Q constrains transmitted symbols to a finite constellation [21], while Hu et al. proposed a masked VQ-VAE framework that represents visual features as discrete codebook indices[22]. These methods provide different ways to represent and transmit image information.

Generative models have also been used to improve image reconstruction at the receiver. Erdemir et al. incorporated generative models into JSCC to improve perceptual quality under limited transmission resources [23]. Wu et al. proposed CDDM, which uses diffusion-based denoising to reduce channel noise [24]. Qiao et al. used pretrained diffusion models with semantic conditioning for generative semantic communication [25], while Zhang et al. incorporated semantic guidance into diffusion-based wireless image reconstruction [26]. In addition, CLIP maps images and text into a shared embedding space [27], while latent diffusion models enable conditional image generation in a compressed latent space [28]. These studies demonstrate the potential of generative models and cross-modal information for image reconstruction.

These methods have also been explored for underwater image communication. Peng et al. incorporated physical priors into underwater image semantic communication [29]. Chen et al. used large language models to understand image semantics and prioritize information, together with conditional diffusion for image recovery [30]. Liang and Ma applied adaptive RVQ to obtain compact visual representations for underwater image transmission [31]. These studies have explored different ways to use physical knowledge, language models, and discrete visual representations in underwater communication.

Overall, previous studies have advanced image representation, digital transmission, generative reconstruction, and underwater semantic communication. However, several issues remain. First, textual descriptions convey the main semantic content of an image but cannot fully represent its spatial structure and fine visual details. Although generative models can infer missing content from learned priors, the generated details may differ from those in the original image when sufficient visual information is unavailable. Second, transmitting supplementary visual information can improve reconstruction, but increases transmission overhead and exposes the additional information to channel errors. Therefore, efficient compression and reliable transmission of visual information remain important. Third, generative reconstruction needs to balance image quality and inference efficiency. Moreover, a visually plausible image does not necessarily preserve the semantics of its corresponding text description. These issues call for a framework that combines cross-modal visual recovery, efficient visual information transmission, and semantically consistent image generation.

To address these issues, we propose a cross-modal semantic successive-refinement framework for generative image communication over underwater acoustic channels. The proposed framework combines textual semantics, compact visual features, and compressed visual side information to progressively reconstruct the transmitted image. First, at the receiver, textual semantics and the received Bottleneck features are used to predict a base estimate of the intermediate visual representation. The RVQ-decoded visual side information is then used for residual correction, further refining the base estimate with additional image-specific visual details. This process improves the recovery of visual information that cannot be fully represented by the text and Bottleneck features alone. Second, RVQ converts high-dimensional visual side information into compact discrete codebook indices to reduce transmission overhead, while unequal error protection (UEP) assigns different protection levels to the transmitted information according to its importance, improving transmission reliability. Finally, the refined intermediate representation and text embedding are fed into a one-step Pixel MeanFlow model [32] to generate the final RGB image, reducing the inference cost associated with iterative generation. During training, a CLIP-based text–image consistency loss [27] is further introduced to encourage semantic alignment between the reconstructed image and its corresponding text description.

The main contributions of this paper are summarized as follows:

**1) We propose a cross-modal semantic successive-refinement framework for underwater image communication.** The receiver first obtains a base estimate of the intermediate visual representation from textual semantics and Bottleneck features, and then refines it using the received visual side information. A one-step Pixel MeanFlow model subsequently reconstructs the final RGB image, enabling cascaded recovery from semantic and visual representations to the pixel domain.

**2) We develop an RVQ-based visual side-information compression and unequal error protection scheme.** RVQ converts high-dimensional visual information into compact discrete codebook indices, reducing the amount of information required for transmission. UEP further allocates different protection levels to transmitted information according to its importance, including different RVQ quantization stages.

**3) We introduce a CLIP-based cross-modal semantic consistency constraint for image reconstruction.** Frozen CLIP image and text encoders are used to measure the semantic similarity between the reconstructed image and its corresponding text description. This training-time constraint encourages the generated image to preserve the textual semantics throughout the cascaded reconstruction process.

## II. General System Model

### *A. Multimodal Representation for Semantic Transmission*

The overall architecture of the proposed generative system is shown in Fig. 1, which is capable for cross-modal data semantic transmission with successive refinement. To start with, an RGB image $X \in R^{3\times H\times W}$ is to be transmitted, let its paired text description be $T$, such textual description can be generated by a well-trained Vision-Language-Model (VLM). At the transmitter, a semantic encoder $E_s(\cdot)$ is leveraged to extract the several levels of visual representations of the input image. In our work, two levels of representations, named of the bottleneck feature $F_B$ and the intermediate feature $F_I$, are extracted:

$$(F_B, F_I) = E_s(X) \tag{1}$$

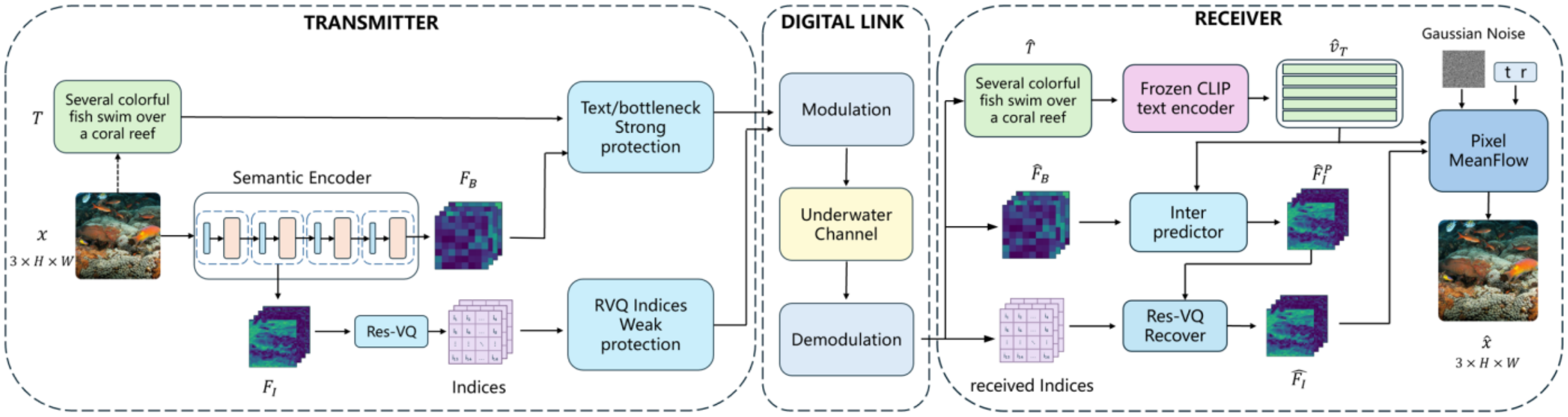

**Fig. 1.** Overall architecture of the cross-modal semantic successive-refinement generative communication system.

Specifically, the semantic encoder is instantiated by the combination of a convolutional neural network and a Swin-Transformer-fashion [33] network. The bottleneck feature is highly compressed and mainly retains abstract information that is irrecognizable, whereas the intermediate feature is less-compressed, which can be further used to recover more details in the reconstruction phase.

To note that, the raw image, the intermediate feature and bottleneck feature form a represenation dependency chain, i.e. intuitively both the intermediate and bottleneck feature are the representation of the raw image, the bottleneck feature can be also regarded as the representation of the intermediate feature. Such chain is useful to illustrate the following hierarchical reconstruction.

The text description $T$ provides cross-modal information about object categories, scene relationships, and overall semantics. These different levels or types of semantic information play distinct but complementary roles during the reconstruction process at the receiver.

During the practical data transmission, the complete intermediate feature contains a large amount of data, which significantly increase the transmission overhead. Therefore, a compression method should be specifically applied to the intermediate feature. Unlike the bottleneck feature, which achieves compression through task-oriented semantic abstraction and may discard fine-grained feature information, the proposed approach adopts a separate feature-level compression pathway (e.g., vector quantization). This pathway directly compresses the intermediate feature for transmission, aiming to reduce communication overhead while preserving its original feature representation as much as possible.

Denoting the compression process of the intermediate feature as $Comp(\cdot)$, the compression result of the massive intermediate information is

$$S = Comp(F_I) \tag{2}$$

where $S$ represents the intermediate compression information, which is actually transmitted over the underwater acoustic link. At the receiver side, the decoder jointly uses textual semantics, bottleneck features as well as the intermediate feature that is additionally recovered from the intermediate compression information $S$ .

*B. Cascaded Reconstruction for Generative Reception*

As mentioned above, the textual description, bottleneck features, and compressed intermediate features are gathered, encoded, and transmitted through the underwater acoustic channel. Conventionally, these heterogeneous information sources are assembled into a single transmission block and protected using the same channel coding scheme, thereby receiving a uniform level of error protection.

However, such a strategy overlooks their distinct roles in receiver-side reconstruction. Specifically, the textual description provides explicit high-level semantic guidance, conveying essential information such as object categories, scene attributes, and semantic relationships. The bottleneck features encode abstract semantic representations and global contextual information, which are crucial for maintaining the overall scene structure and semantic consistency. In contrast, the intermediate features preserve richer spatial information and fine-grained details, such as object boundaries, local structures, and textures, complementing the semantic abstraction of the bottleneck features.

Consequently, transmission errors in these three information streams may lead to different types and degrees of reconstruction degradation. This motivates the adoption of unequal error protection (UEP), whereby different levels of channel protection are assigned to each information stream according to its contribution and sensitivity to reconstruction quality. Each stream is then quantized and transmitted to the receiver through a digital communication link with an appropriately configured error-protection level. The specific underwater acoustic channel configuration is introduced in the experimental settings in Section IV.

To elaborate, the receiver performs image reconstruction through two cascaded generative stages. First, the received text description $\hat{T}$ is converted into a text embedding $\hat{v}_T$ by a frozen CLIP Text Encoder [27]. Together with the recovered bottleneck feature, this embedding is fed into the feature reconstruction network, named as the Inter Predictor $P_\theta$, to obtain an initial prediction of the visual-side intermediate representation:

$$\hat{F}_I^P = P_\theta(\hat{v}_T, \hat{F}_B) \in R^{C_I \times H_I \times W_I} \tag{3}$$

where $\hat{F}_B$ denotes the received bottleneck feature after the acoustic channel. $P_\theta(\cdot)$ denotes the feature-domain Inter Predictor.

As the predicted intermediate feature $\widehat{F}_I^P$ is obtained, the receiver uses the recovered visual-side information to correct the prediction and obtain a more enhanced intermediate visual representation. This process of intermediate feature enhancement can be expressed in a compact form as

$$\widehat{F}_I = F_{\psi}\left(\widehat{F}_I^P, \hat{S}\right) \in R^{C_I \times H_I \times W_I} \quad (4)$$

where $\hat{S}$ denotes the intermediate compression information recovered from the acoustic channel and $F_{\psi}(\cdot)$ descripts the procedure of a CNN-based intermediate feature recovery and a subsequent fusion operation.

To elaborate, the Equation (4) is envisioned by the RVQ scheme that is mainly adopted in this paper – the received discrete indices are used to select the subset of the shared codebook with transmitter, then obtain the recovered intermediate compression information $\hat{S}$ by aggregating all the selected codewords. The predicted intermediate representation $\widehat{F}_I^P$ is further enhanced to $\widehat{F}_I$ with the assistance of $\hat{S}$.

After the restored enhanced intermediate representation $\widehat{F}_I$ has been obtained, in the second stage, a one-step Pixel MeanFlow generator reconstructs the image in pixel-domain. The recovered intermediate representation and text embedding are jointly leveraged as generation conditions, yielding the reconstructed image:

$$\hat{X} \sim P_{\omega}\left(X \middle| \hat{v}_T, \widehat{F}_I\right) \quad (5)$$

where $P_{\omega}\left(X \middle| \hat{v}_T, \widehat{F}_I\right)$ denotes the conditional image distribution modeled by the one-step Pixel MeanFlow generator. Through the two-stage cascaded reconstruction process, the receiver first recovers a richer intermediate representation from compact cross-modal semantic conditions and then reconstructs a pixel-domain image from this representation and, consequently, achieving semantic successive refinement across representation domains.

Additionally, to maintain semantic consistency between the progressively reconstructed image and its corresponding textual description, a cross-modal semantic consistency loss is introduced. Such loss encourages the reconstructed visual content to preserve the high-level semantic information conveyed by the transmitted text, thereby promoting semantic alignment across modalities. Specifically, the reconstructed image $\hat{X}$ and its corresponding received textual description $\hat{T}$ are fed into the image encoder and text encoder of a pretrained CLIP model as

$$\begin{cases} \hat{v}_I = CLIP_I(\hat{X}) \\ \hat{v}_T = CLIP_T(\hat{T}) \end{cases} \quad (6)$$

respectively, the resulting image and text embeddings are mapped into a shared semantic space, where their cosine similarity is calculated to quantify the cross-modal semantic consistency,

$$\mathcal{L}_{CLIP} = 1 - \hat{v}_I{}^T \hat{v}_T \quad (7)$$

which is introduced to the training process of the multimodal semantic receiver.

## III. Implementation Details

The previous section presents a general formulation of the proposed system end-to-end. In this section, the details of the realization are elaborated.

### *A. Hybrid Source and Channel Coding*

To start with, we introduce the transmission procedure. In conventional communication systems, the data to be transmitted are first processed by a source encoder, which removes statistical redundancy to reduce the data volume, such as JPEG for image compression and H.264 or H.265 for video compression. The resulting compressed bitstream is then fed into a channel encoder, which introduces redundancy to improve transmission reliability over noisy channels. For example, Polar codes are adopted for control-channel coding, while LDPC codes are employed for data-channel coding in 5G New Radio (NR).

In contrast to the conventional paradigm, our transmitter adopts a hybrid approach to multimodal information representation and transmission. Specifically, it takes an image $X$ and its corresponding textual description $T$, generated by a pretrained vision-language model (VLM), as inputs. For the image modality, the semantic encoder $E_s$ extracts two feature representations at different abstraction levels, namely, the bottleneck feature $F_B$ and the intermediate feature $F_I$, as defined in (1). To reduce the transmission overhead associated with the intermediate feature, $F_I$ is further compressed to obtain a compact representation $S$. For the textual modality, the original text $T$ is retained as the source representation and directly used for transmission, without additional semantic feature extraction.

Once these multimodal representations, i.e., $F_B$, $S$, and $T$, are obtained, they are independently fed into dedicated channel encoders. Since the three representations play distinct roles in receiver-side reconstruction and exhibit different sensitivities to transmission errors, different levels of channel protection are assigned to each stream. This enables the transmitter to allocate channel-coding redundancy according to the respective importance of each information stream.

After the source representation and bitstream preparation, the three resulting bitstreams corresponding to $F_B$, $S$, and $T$ are separately fed into three independent LDPC encoders, each configured with a different code rate. Specifically, let $K_j$ and $N_j$ denote the input and output bit lengths of the j-th stream, respectively. Its effective code rate is defined as

$$\mathcal{R}_j = K_j / N_j, \qquad j \in B, S, T. \quad (8)$$

By assigning different code rates to the three streams according to their respective importance and sensitivity to transmission errors, unequal error protection (UEP) is achieved. In particular, a lower code rate introduces more redundancy and generally provides stronger error protection. The three channel-coded bitstreams are then concatenated in a predefined order to form a single coded bitstream:

$$b = b_B || b_S || b_T \quad (9)$$

where $b_B$, $b_S$, and $b_T$ denote the LDPC-coded bitstreams, respectively, and || represents bitstream concatenation. Finally, the concatenated bitstream is mapped onto QPSK symbols, with

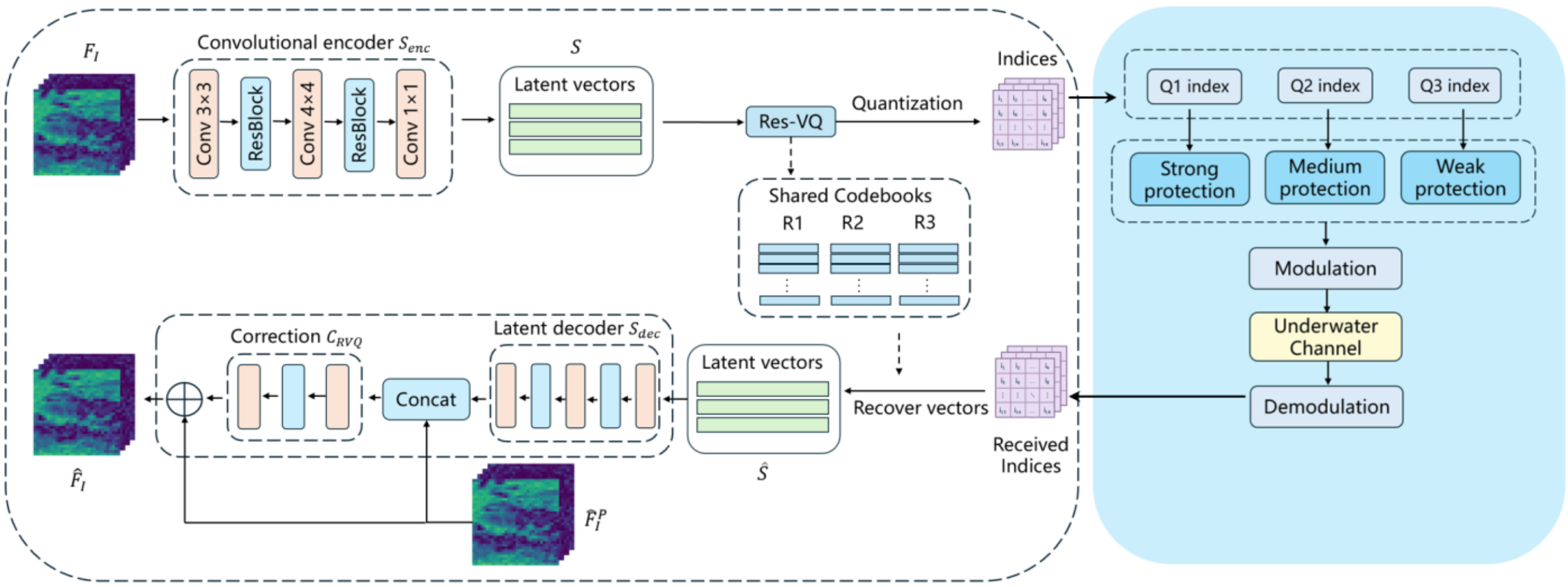

**Fig. 2.** RVQ-based Intermediate information transmission.

every two consecutive coded bits forming one modulation symbol, yielding the transmitted symbol sequence:

$$\phi_m = \mu_{\text{QPSK}}(b_{2m-1}, b_{2m}) \tag{10}$$

Such symbols are furtherly modulated with OFDM scheme and transmitted into the noisy acoustic channel environment.

*B. Base Prediction of Intermediate Feature Reconstruction*

At the receiver, the received acoustic signal undergoes the corresponding reverse processing procedure, including QPSK demodulation, bitstream separation, and LDPC decoding. This yields three recovered multimodal representations, namely, the bottleneck feature $\hat{F}_B$, the intermediate feature indices $\hat{S}$, and the textual description $\hat{T}$ . Based on these decoded representations, the receiver proceeds with the reconstruction process.

In the first reconstruction branch, the receiver generates a base prediction of the intermediate feature $\hat{F}_I^P$ by leveraging the decoded bottleneck feature $\hat{F}_B$ and the received textual description $\hat{T}$, as formulated in (3). Specifically, to incorporate textual information as a semantic condition, $\hat{T}$ is first fed into a frozen CLIP text encoder to obtain a global semantic embedding the same as (6)

$$\hat{v}_T = CLIP_T(\hat{T}) \tag{11}$$

This embedding is then transformed into a set of text-conditioning tokens through a learnable multilayer perceptron (MLP).

$$\hat{C}_T = MLP_\eta(\hat{v}_T) \tag{12}$$

The text-conditioning tokens provide high-level semantic cues, including object categories and scene relationships.

In parallel, the bottleneck feature $\hat{F}_B$ is processed by a CNN-based projection network $G_\phi$ to generate visual-conditioning tokens while preserving its spatial layout,

$$\hat{V}_B = G_\phi(\hat{F}_B) \tag{13}$$

These tokens provide complementary visual structural information, such as spatial organization and coarse structural characteristics. By jointly exploiting the semantic guidance from the text and the structural priors encoded in the bottleneck feature, the Inter Predictor generates a base prediction of the intermediate feature, which serves as the initial representation for subsequent residual refinement.

The Inter Predictor is implemented as a conditional prediction network consisting of eight stacked Transformer blocks. Specifically, the text-conditioning tokens and bottleneck-conditioning tokens are first concatenated along the token dimension to form a unified multimodal sequence:

$$Z_0 = [\hat{C}_T; \hat{V}_B] \tag{14}$$

where $\hat{C}_T$ and $\hat{V}_B$ denote the text-conditioning and visual-conditioning tokens, respectively. The concatenated sequence is then processed by the Transformer blocks, where multi-head self-attention enables interactions between textual and visual representations, facilitating the fusion of high-level semantic cues and spatial structural information. The resulting multimodal representation is subsequently passed through an output projection layer to generate the predicted intermediate feature:

$$\hat{F}_I^P = G_{out}\left(T_\theta^{(8)}(Z_0)\right) \tag{15}$$

where $T_\theta^{(8)}$ denotes the stack of eight Transformer blocks and $G_{out}$ represents the output projection.

During training, the ground-truth intermediate feature $F_I$, extracted by the transmitter-side semantic encoder, serves as the supervision target. To ensure consistent feature scaling, the prediction and ground-truth feature are normalized before calculating the mean squared error (MSE) loss:

$$\mathcal{L}_{\text{pred}} = \frac{1}{D_I}\left\|N(\hat{F}_I^P) - N(F_I)\right\|_F^2 \tag{16}$$

where $N(\cdot)$ denotes the feature normalization operation and $D_I$ is the total number of elements in the intermediate feature. By minimizing this loss, the Inter Predictor learns to reconstruct the intermediate visual representation from the compact multimodal conditioning information.

After training, the Inter Predictor is frozen to generate a base estimate of the intermediate feature from the decoded bottleneck feature and textual description. When only these two sources are transmitted, the base estimate provides visual conditioning for downstream image reconstruction. When RVQ indices are additionally received, the decoded residual features are used to further refine the estimate. By maintaining a fixed prediction baseline, the frozen Inter Predictor provides a stable reference for the subsequent residual refinement and feature fusion.

### C. RVQ-based Intermediate Feature Residue Prediction

The frozen Inter Predictor described in the previous subsection generates a base estimate of the intermediate feature $\hat{F}_I^P$ from the received bottleneck feature $\hat{F}_B$ and textual description $\hat{T}$. Although this estimate approximates the ground-truth intermediate feature $F_I$, fine-grained visual details cannot be fully recovered from these compact conditions alone. Therefore, additional visual side information is transmitted and used to learn the residue part of $F_I$, in order to refine the base estimate.

To elaborate this residue branch, specifically, let's revisit equation (2) in the transmission part, $F_I$ in compressed by a CNN-based encoder $S_{enc}$, as illustrated in Fig. 2. This encoder reduces its spatial resolution and channel dimensionality while preserving information relevant to receiver-side refinement. It is trained end-to-end to retain the visual details that cannot be adequately recovered by the Inter Predictor. The encoder outputs a compressed feature tensor

$$S_{raw} \in R^{C_S \times H_S \times W_S} \tag{17}$$

Before quantization, $S_{raw}$ is reshaped into a matrix by flattening its spatial dimensions:

$$S = reshape(S_{raw})^T \in R^{H_S W_S \times C_S} \tag{18}$$

Consequently, each row of $S$ represents a $C$-dimensional latent vector, resulting in $H_S W_S$ latent vectors in total. These vectors are then quantized using a multistage Residual Vector Quantizer (RVQ)[34]. Specifically, at each quantization stage, the residuals of all latent vectors are quantized using a shared codebook. Different quantization stages employ distinct codebooks, while all codewords within a codebook have the same dimensionality $C_S$ as the latent vectors. The resulting discrete codebook indices represent the quantized form of $S$ and are transmitted over the digital link. By varying the number of quantization stages, RVQ supports different trade-offs between transmission overhead and reconstruction fidelity.

RVQ uses multiple quantization stages to encode the residual error that remains insufficiently represented by the preceding stage. First, the original latent representation is used as the initial quantization residual, i.e., $r_0 = S$ . At the $l$ -th quantization stage, quantizer $Q(r; R)$ selects the nearest codeword from its codebook $R$ according to the current residual $r$, and the selected codeword is denoted by $q$. The portion already represented by this codeword is then subtracted from the current residual:

$$[q_l, i_l] = Q(r_{l-1}, R_l) \tag{19a}$$

$$r_l = r_{l-1} - q_l = r_{l-1} - R_l\{i_l\} \tag{19b}$$

where $q_l$ represents the codeword obtained at quantization stage $l$, $i_l$ denotes the index of $q_l$ in the $l$-th level codebook $R_l$ and $r_l$ denotes the residual error that remains unrepresented after this stage. Thus, the first stage provides a coarse approximation of the original latent feature, while subsequent stages continue to represent visual information not adequately captured by the preceding stages. What is actually sent through the digital link is not the codeword vectors $q_l$ themselves, but the codebook indices $i_l$ selected by quantizer of each stage, thereby substantially reducing the amount of transmitted visual side information.

At the reception part, with $L$ RVQ stages, the receiver retrieves the corresponding indices of codewords from the shared codebooks with transmitter using the received indices and summates them to obtain the recovered latent visual representation:

$$\hat{S} = \sum_{l=1}^{L} R_l\{\hat{\imath}_l\} \tag{20}$$

where $\hat{\imath}_l$ denotes the received index at stage $l$. It can be seen that, due to the erroneous channel, the codeword indices may deviate from their correct values and in turn, causes the error latent vectors. To solve this, a learnable error-correction network is inserted to mitigate the effects from the codeword indices error.

In particular, as shown in Fig. 2 the erroneous $\hat{S}$ is firstly decompressed by another CNN-based decoder $S_{dec}$, which is a reverse procedure of $S_{enc}$. The output of the decoder is

$$\hat{F}_I^{dec} = S_{dec}(\hat{S}) \in R^{C_{dec} \times H_I \times W_I} \tag{21}$$

whose dimension is slightly different from $\hat{F}_I \in R^{C_I \times H_I \times W_I}$. Then, the $\hat{F}_I^P$ and $\hat{F}_I^{dec}$ are concatenated as the input of an error-correction network $C_{RVQ}$ to predict the difference between the ground truth intermediate feature $F_I$ and the base-estimated one $\hat{F}_I^P$. Finally, they are summed up to generate a more accurate intermediate feature

$$\hat{F}_I = \hat{F}_I^P + C_{\text{RVQ}}\left(\text{Concat}\left[\hat{F}_I^{dec}, \hat{F}_I^P\right]\right) \tag{22}$$

which is the detail implementation of equation (4).

Based on the base estimate $\hat{F}_I^P$ generated by the frozen Inter Predictor (Eq. (15)), we further train the RVQ-based residual refinement branch. During this process, the Inter Predictor remains frozen, while the original intermediate feature $F_I$ serves as the reconstruction target.

To enable effective reconstruction with different numbers of RVQ stages, a stage-wise reconstruction supervision strategy is applied. Specifically, for a three-stage RVQ configuration, three reconstruction cases are considered, using the cumulative quantized representations from the first stage, the first two stages, and all three stages, respectively. The corresponding latent representations are given by

$$\hat{S}^{(k)} = \sum_{l=1}^{k} R_l\{\hat{\imath}_l\}, \quad k \in \{1,2,3\} \tag{23}$$

Each cumulative representation is processed by the same CNN-based decoder $S_{dec}$ and the same error-correction network $C_{RVQ}$. Together with the fixed base estimate $\hat{F}_I^P$, the decoded feature is used to generate an enhanced intermediate feature:

$$\hat{F}_I^{(k)} = \hat{F}_I^P + C_{RVQ}\left(\text{Concat}\left[S_{\text{dec}}(\hat{S}^{(k)}), \hat{F}_I^P\right]\right) \tag{24}$$

Accordingly, three reconstructed features, $\hat{F}_I^{(1)}$, $\hat{F}_I^{(2)}$, and $\hat{F}_I^{(3)}$, are obtained and individually supervised against the ground-truth intermediate feature $F_I$. The overall reconstruction loss is defined as the average of the three stage-wise reconstruction losses:

$$\mathcal{L}_{\text{rec}} = \frac{1}{3}\sum_{k=1}^{3} \text{MSE}\left(\hat{F}_I^{(k)}, F_I\right) \tag{25}$$

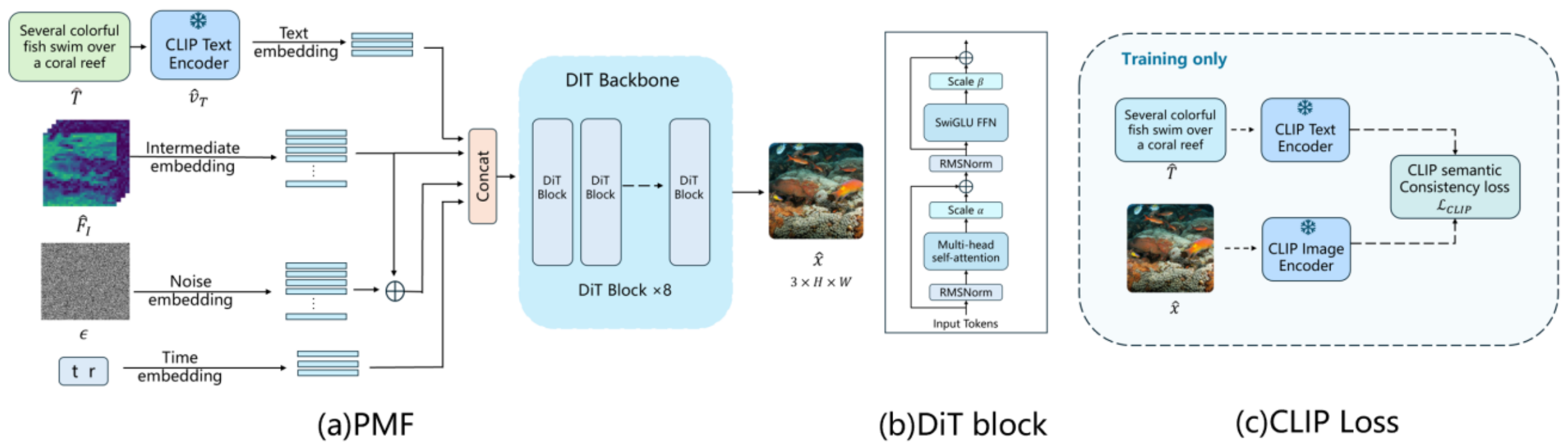

**Fig. 3.** Image reconstruction using Pixel MeanFlow.

By jointly optimizing these three reconstruction cases, the residual refinement branch is explicitly trained to exploit different levels of RVQ information. In particular, the first-stage supervision ensures that even a single quantization stage can provide useful residual correction, while subsequent stages progressively improve the reconstruction by incorporating additional residual information. Consequently, the convolutional encoder $S_{enc}$, latent decoder $S_{dec}$, and correction network $C_{RVQ}$ are optimized jointly, and the whole residue branch network is encouraged to maintain effective reconstruction performance at different RVQ depths, rather than relying solely on the complete three-stage representation.

In addition to the reconstruction loss, a commitment loss is introduced to encourage the input to each quantizer to remain close to its selected codewords. For the three-stage model, this loss and the combined training objective are expressed as

$$\mathcal{L}_{com} = \frac{1}{3}\sum_{l=1}^{3} MSE\big(r_{l-1}, sg(q_l)\big) \tag{26a}$$

$$\mathcal{L}_{RVQ} = \mathcal{L}_{rec} + \lambda_{com}\mathcal{L}_{com} \tag{26b}$$

respectively, where $r_{l-1}$ is the input residual at stage $l$, with $r_0 = S$, and $sg(\cdot)$ denotes the stop-gradient operation [31], which preserves its argument during forward computation but prevents gradients from propagating through it. The coefficient $\lambda_{com}$ controls the weight of the commitment loss. In summary, these objectives encourage the encoder to produce latent representations that support reconstruction at different refinement levels while remaining amenable to quantization.

The residual branch network is optimized through backpropagation. Since RVQ involves discrete codeword selection, a straight-through estimator (STE) is employed to enable reconstruction gradients to propagate through the quantization operation to the encoder. Meanwhile, the codebooks are initialized by sampling latent vectors for the first stage and residual vectors for subsequent stages. During training, they are updated using an exponential moving average (EMA), which gradually moves each codeword toward the centroid of its assigned vectors. In this way, the encoder learns latent representations suitable for reconstruction and quantization, while the codebooks progressively adapt to the evolving latent and residual distributions. After training, the codebooks are fixed and shared between the transmitter and receiver.

### *D. Low-Rank Intermediate Representations Based on Tucker and CP*

As alternative non-learned representations to the RVQ-based approach, we further consider two low-rank tensor decomposition methods, namely, Tucker decomposition and CP decomposition [35]. Unlike RVQ, which relies on training data to learn the encoder, codebooks, and decoder, these methods directly approximate the intermediate feature tensor through low-rank decomposition, without requiring offline model training or an additional training dataset. They provide an alternative approach to visual side-information compression, enabling a comparative investigation of learned discrete representations and non-learned low-rank representations.

Given an intermediate feature tensor $F_I$ , Tucker decomposition approximates it using a compact core tensor and a set of factor matrices:

$$F_I \approx G \times_1 U^{(1)} \times_2 U^{(2)} \times_3 U^{(3)} \tag{27}$$

where $G$ denotes the core tensor, and $U^{(1)}$, $U^{(2)}$, and $U^{(3)}$ are factor matrices capturing the low-rank subspaces along the respective tensor modes. By restricting the decomposition rank of each mode, Tucker decomposition exploits inter-dimensional correlations to reduce the representation size. The original feature can subsequently be reconstructed from the core tensor and factor matrices at the receiver.

CP decomposition provides another low-rank approximation by expressing the original tensor as a sum of rank-one components:

$$F_I \approx \textstyle\sum_{r=1}^{R} \lambda_r a_r \circ b_r \circ c_r \tag{28}$$

where $R$ denotes the CP rank, $a_r$, $b_r$, and $c_r$ are factor vectors along the three tensor modes, and $\lambda_r$ is the weight of the $r$th rank-one component, and $\circ$ denotes the outer product. Unlike Tucker decomposition, CP does not require a separate core tensor, resulting in a more compact factorized representation. However, its approximation performance and numerical stability are sensitive to the selected rank and the decomposition procedure.

At the receiver, the recovered low-rank representations from Tucker or CP decomposition serve as additional visual side information. They are fused with the base intermediate feature prediction from the cross-modal predictor through cross-attention, providing complementary visual information for subsequent reconstruction. These methods, together with RVQ, allow us to investigate the differences between learned discrete residual quantization and non-learned continuous low-rank decomposition for visual feature transmission over underwater acoustic channels.

### *E. Pixel MeanFlow RGB Reconstruction and Cross-Modal Semantic Consistency Loop*

With the recovered intermediate feature $\hat{F}_I$ and text embedding $\hat{v}_T$, the receiver employs Pixel MeanFlow model [32] to reconstruct the RGB image in a single network forward pass. The generation process is formulated through two complementary representations of the image state $X_t$.

The first representation defines a linear interpolation between a ground-truth image $X$ and an independent Gaussian noise sample $\epsilon$:

$$X_t = (1-t)X + t\epsilon, \quad \epsilon \sim \mathcal{N}(0, I), \quad t \in [0,1] \tag{29}$$

Here, $t = 0$ corresponds to the clean image, whereas $t = 1$ corresponds to pure noise. Since both $X$ and $\epsilon$ are known during training, differentiating the interpolation yields the ground-truth instantaneous velocity:

$$v_{\text{target}} = \frac{dX_t}{dt} = \epsilon - X \tag{30}$$

The second representation describes the same time-indexed state in terms of the displacement along a generative trajectory conditioned on the recovered intermediate feature $\hat{F}_I$ and text embedding $\hat{v}_T$:

$$X_t = X_r + (t-r)u\big(X_t, r, t; \hat{F}_I, \hat{v}_T\big), \quad 0 \le r < t \le 1 \tag{31}$$

Here, $X_r$ denotes the image state at the earlier target time $r$, while $u\big(X_t, r, t; \hat{F}_I, \hat{v}_T\big) = (X_t - X_r)/(t-r)$ represents the average velocity over the interval $[r, t]$. Unlike the linear interpolation in (29), this representation characterizes the net displacement along the conditional generative trajectory. Since the target state $X_r$ is unavailable during generation, the average velocity is estimated by a network $u_\omega\big(X_t, r, t; \hat{F}_I, \hat{v}_T\big)$, which uses $\hat{F}_I$ and $\hat{v}_T$ as visual and semantic conditions, respectively.

Differentiating (31) with respect to $t$, while holding $r$ and its corresponding trajectory state fixed, with the conditioning variables $\hat{F}_I$ and $\hat{v}_T$ also held fixed, relates the estimated average velocity to the instantaneous velocity:

$$V_\omega\big(X_t, r, t; \hat{F}_I, \hat{v}_T\big) = u_\omega + (t-r)\frac{du_\omega}{dt} \tag{32}$$

The derivative is the total derivative along the time-dependent state $X_t$, with $\hat{F}_I$ and $\hat{v}_T$ treated as fixed conditions. This relationship enables the average-velocity network to be trained using the ground-truth instantaneous velocity from (30):

$$\mathcal{L}_{PMF} = \mathbb{E}_{X,\epsilon,\hat{F}_I,\hat{v}_T,r,t}\left[\left\|V_\omega\big(X_t, r, t; \hat{F}_I, \hat{v}_T\big) - v_{\text{target}}\right\|_2^2\right] \tag{33}$$

Pixel MeanFlow parameterizes the average velocity through an image-valued network output $\tilde{X}_\omega\big(X_t, r, t; \hat{F}_I, \hat{v}_T\big)$, which jointly depends on the noisy image state, time interval, recovered visual feature, and text embedding. The corresponding target-state prediction is

$$u_\omega\big(X_t, r, t; \hat{F}_I, \hat{v}_T\big) = \frac{X_t - \tilde{X}_\omega\big(X_t, r, t; \hat{F}_I, \hat{v}_T\big)}{t} \tag{34a}$$

$$\hat{X}_r = X_t - (t-r)u_\omega = \frac{r}{t}X_t + \left(1 - \frac{r}{t}\right)\tilde{X}_\omega \tag{34b}$$

For one-step reconstruction, setting $t = 1, r = 0$, and $X_1 = \epsilon$ gives

$$\hat{X}_0 = \tilde{X}_\omega\big(\epsilon, 0,1; \hat{F}_I, \hat{v}_T\big) \tag{35}$$

As illustrated in Fig. 3, the Pixel MeanFlow network incorporates the recovered intermediate feature $\hat{F}_I$ and text embedding $\hat{v}_T$ as complementary visual and semantic conditions. The noisy image, intermediate feature, text embedding, and time information are projected into corresponding tokens and jointly processed by Transformer blocks, enabling the image tokens to attend to both the visual structure provided by $\hat{F}_I$ and the semantic information conveyed by $\hat{v}_T$. Additionally, spatially aligned intermediate features derived from $\hat{F}_I$ are added to the corresponding image tokens to preserve spatial correspondence. The resulting image tokens are projected and rearranged to form the conditional image-valued output $\tilde{X}_\omega\big(X_t, r, t; \hat{F}_I, \hat{v}_T\big)$.

To further improve reconstruction fidelity and semantic consistency, L1 and LPIPS losses are employed to constrain pixel-level and perceptual discrepancies, while a CLIP-based loss encourages agreement between the reconstructed image and its textual description. The overall training objective is

$$\mathcal{L}_{\text{train}} = \mathcal{L}_{\text{PMF}} + \lambda_{LPIPS}\mathcal{L}_{\text{LPIPS}} + \lambda_{L1}\mathcal{L}_{L1} + \lambda_{CLIP}\mathcal{L}_{\text{CLIP}} \tag{36}$$

The velocity derivative and auxiliary losses are evaluated only during training. At inference, the recovered intermediate feature $\hat{F}_I$ and text embedding $\hat{v}_T$ guide the trained Pixel MeanFlow network to reconstruct the final RGB image directly from Gaussian noise in a single forward pass, with the visual and semantic conditions jointly determining the reconstructed image content.

## IV. Experimental Results And Analysis

### *A. Experimental Settings*

In our work, a subset of 50,000 images is selected from the COCO 2017 dataset [36], with 40,000 images used for training, 5,000 for validation, and 5,000 for testing. Each image is associated with five manually annotated natural-language descriptions, and the corresponding official captions are used as text descriptions throughout the training, validation, and testing stages. In the experiments, one official caption is selected for each image as its text condition, and all images are uniformly preprocessed into RGB images of size $128 \times 128$. For the underwater channel, we use the NOF1 measured channel records from the Watermark V1 dataset [37]. The Watermark dataset contains underwater acoustic channel measurements collected in various representative marine environments and reflects channel variations under different propagation distances, water depths, frequency bands, and multipath conditions. The NOF1 record selected in this work was collected in a shallow fjord environment with relatively fixed transmitter and receiving-hydrophone positions. Its sounding signal has a duration of 32.9 s, an operating frequency band of 10–18 kHz, a propagation distance of approximately 750 m, and a water depth of approximately 10 m. The record was repeatedly collected over an observation period of about 33 min and can describe the time-varying characteristics of the underwater acoustic channel.

The system consists of a semantic encoder, an Inter Predictor, a side-information compression module, and a Pixel MeanFlow image generator. The semantic encoder combines a CNN and a Swin Transformer and outputs a bottleneck feature of size 16 × 8 × 8 and an intermediate feature of size 96 × 32 × 32. The RVQ encoder maps the intermediate feature to a latent space of size 64 × 16 × 16, after which multistage residual quantization produces discrete codebook indices. Tucker and CP decompositions are also considered as the alternative training-free low-rank visual side-information representations. Both the Inter Predictor and Pixel MeanFlow employ conditional DiT architectures with a hidden dimension of 384 and eight Transformer blocks. The principal network architectures and experimental parameters are summarized in Table I. All experiments were conducted on a server equipped with an NVIDIA GeForce RTX 3090 GPU and an Intel Xeon Gold 6240 CPU. The time of the inference for a single image, (i.e., including the end-to-end encoding, transmission, reception and decoding, but excluding the period of the underwater propagation and hardware-based channel coding), varies from 40-100 ms, which is practical for real-world deployment.

Table I. Main network architectures and experimental settings

| Module | Configuration |
|---|---|
| Semantic encoder | CNN-Swin; four stages with channels [96, 192, 256, 192]; depths [2, 2, 6, 2]; window size 4 |
| bottleneck feature | 16 × 8 × 8 |
| intermediate feature | 96 × 32 × 32 |
| Inter Predictor | Conditional DiT; hidden dimension 384; eight blocks; 2 × 2 intermediate patch |
| RVQ latent vectors | 64 × 16 × 16 |
| RVQ codebook size | 64, 128, 256, 512 |
| RVQ depth | 1, 2, 3 |
| Tucker ranks | (24, 10, 10), (24, 16, 16) |
| CP ranks | R = 33, R = 59 |
| Pixel MeanFlow generator | Conditional DiT; hidden dimension 384; eight blocks; 8 × 8 RGB patch |

## *B. Effects of RVQ Depth and Codebook Size*

To investigate the effects of RVQ quantization depth and codebook size on reconstruction performance, we vary these two parameters while maintaining a fixed channel SNR of 10 dB, and evaluate the resulting image reconstruction quality.

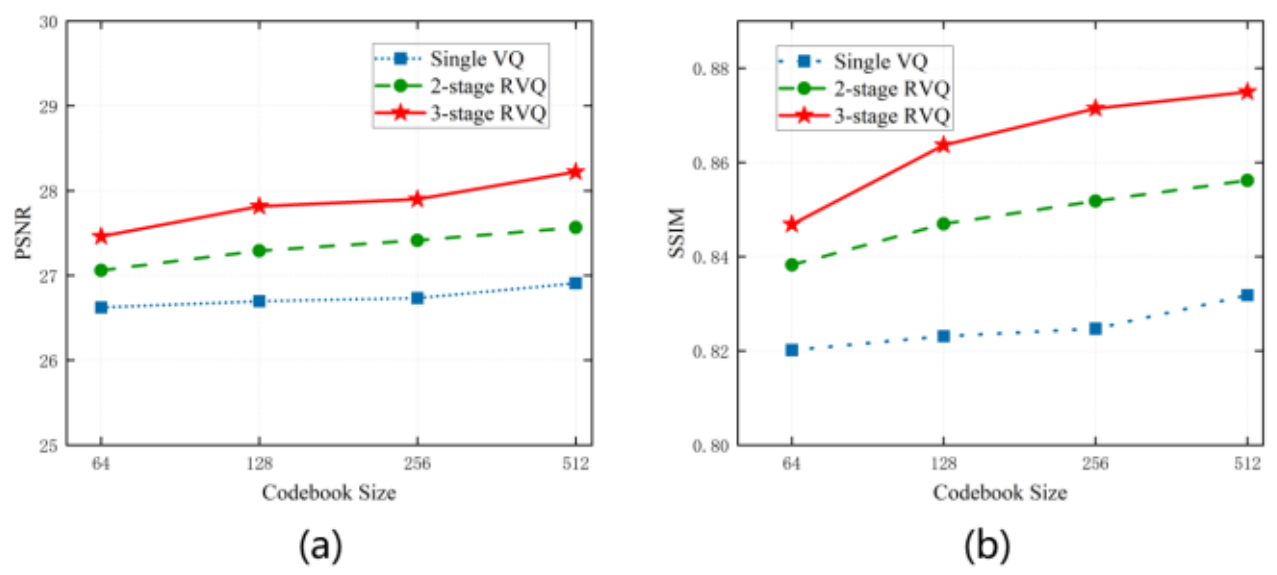

**Fig. 4.** Reconstruction performance for different RVQ codebook sizes and quantization depths at 10 dB channel SNR.

The channel coding rates for the bottleneck and text streams are both fixed at 1/3, whereas all RVQ stages use a code rate of 1/2, ensuring a consistent channel coding rate across different RVQ configurations. Since the bottleneck and text representations contribute to multiple stages of the reconstruction process, they carry more critical information and are therefore assigned lower code rates to provide stronger error protection. As shown in Fig. 4, increasing the RVQ quantization depth leads to consistent improvements in both PSNR and SSIM, indicating that subsequent residual quantization stages capture complementary visual information that remains inadequately represented by preceding stages. As shown in Fig. 4(a), at an SNR of 10 dB and a codebook size of 256, increasing the RVQ depth from one to two and three stages raises the PSNR from 26.7 to 27.4 and 27.9 dB, respectively. However, these improvements come at the cost of increased transmission overhead, as additional quantization stages require transmitting more codebook indices. Similarly, larger codebooks size generally require more bits to represent each index. Fig. 4(a) also shows that, for three-stage RVQ at an SNR of 10 dB, increasing the codebook size from 64 to 128, 256, and 512 yields PSNR values of 27.4, 27.8, 27.9, and 28.2 dB, respectively. These results highlight the trade-off between reconstruction quality and transmission efficiency.

## *C. Comparison of Tucker and CP Tensor-Decomposition Representations*

Tucker and CP tensor decompositions (TD) are leveraged as the alternative training-free visual side-information representations for comparison. As shown in Fig. 5, the higher channel SNR, the better reconstruction performance for both types of low-rank representation, indicating that more reliable channel transmission enables an increased TD representation capacity to provide the receiver with more intact supplementary visual information. As shown in Fig. 5(a), when the channel SNR increases from 5 to 10 dB, the PSNR of Tucker decomposition with ranks (24, 16, 16) increases from 25.9 to 27.4 dB, while that of CP decomposition with rank 59 increases from 25.8 to 28.4 dB. In addition, for both Tucker and CP TD, higher-rank ones also promote the reconstruction performance due to the preservation of the original intermediate feature information, yet indicating higher demands on transmission overhead. The rank comparison in Fig. 5(a) shows that, at an SNR of 10 dB, increasing the Tucker ranks from (24, 10, 10) to (24, 16, 16) improves the PSNR from 26.4 to 27.4 dB. Similarly, increasing the CP rank from 33 to 59 improves the PSNR from 26.9 to 28.4 dB.

From the perspective of representation volume, the visual side information decomposition for Tucker and CP directly consists of a core tensor, factor matrices, or factor vectors, which is substantially larger than those of RVQ, i.e., only the transmission of several discrete indices are needed for the latter. For a fair comparison, Tucker and CP TD take advantages in eliminating model training and reducing dependence on the training distribution, whereas their relatively large representation volume limits the application in resource-constrained underwater acoustic links.

## *D. Ablation Study*

To evaluate the contributions of different information branches and the effectiveness of unequal error protection in cascaded reconstruction, we consider five transmission strategies: “Bneck+Text”, “Text+RVQ”, “Bneck+RVQ”,

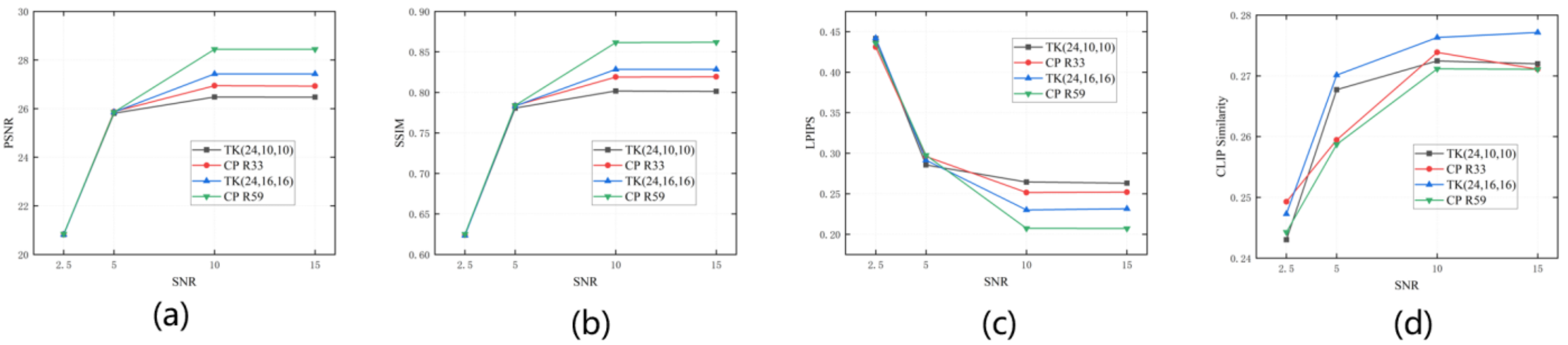

**Fig. 5.** Image reconstruction performance of Tucker and CP low-rank visual side-information representations at different SNR.

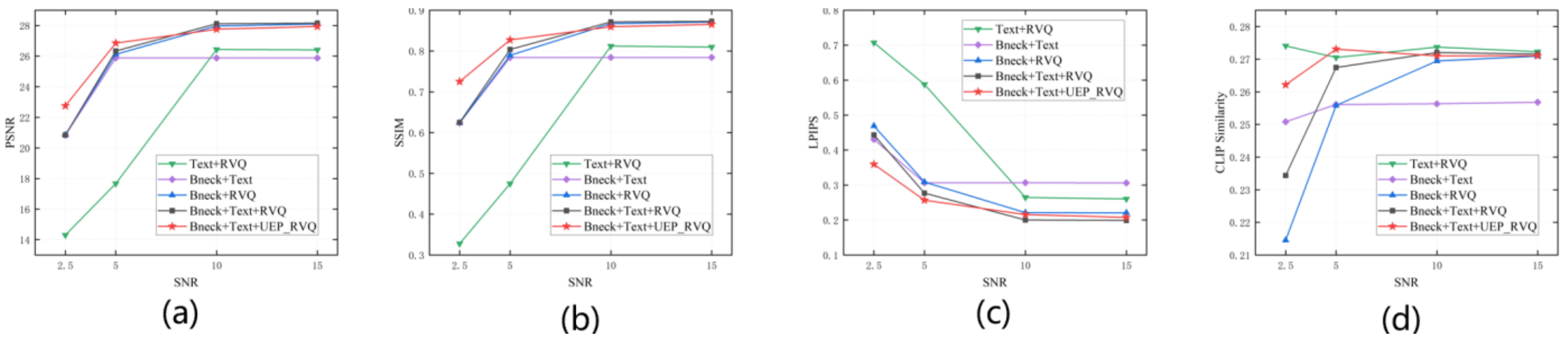

**Fig. 6.** Ablation results for different information branches at various SNR.

"Bneck+Text+RVQ", and "Bneck+Text+UEP_RVQ". Here, "Bneck" denotes the bottleneck feature, "Text" denotes the textual stream, and "RVQ" represents the transmission of indices generated by Residual Vector Quantization. "UEP_RVQ" further indicates that the RVQ indices at different quantization stages are assigned unequal channel protection levels. The first three configurations serve as ablation studies of the information branches: "Bneck+Text" excludes the RVQ-based visual side information, "Text+RVQ" removes the bottleneck branch, and "Bneck+RVQ" removes the text branch. The last two configurations retain all three information branches but differ in their channel protection strategies

Regarding channel coding, the "Bneck" and "Text" streams are consistently assigned a code rate of 1/3 whenever they are included. For conventional RVQ transmission, all three stages use an equal code rate of 1/2. In contrast, "UEP_RVQ" adopts stage-dependent code rates of 1/3, 4/7, and 4/5 for the first, second, and third quantization stages, respectively. Since a lower code rate introduces more redundancy and provides stronger error protection, the first RVQ stage receives the highest protection level, followed by the second and third stages. The experimental results are presented in Fig. 6.

Compared with the "Bneck+Text" strategy, incorporating RVQ ("Bneck+Text+RVQ") improves reconstruction quality, particularly in the higher-SNR region. As shown in Fig. 6(a), at an SNR of 10 dB, "Bneck+Text" achieves a PSNR of 25.8 dB, whereas "Bneck+Text+RVQ" achieves 28.1 dB, corresponding to a gain of 2.3 dB from the additional RVQ side information. This demonstrates that the transmitted visual side information provides complementary details that cannot be fully recovered from the compact cross-modal conditions alone. The qualitative results in Fig. 7 further support this observation. At 10 dB, "Bneck+Text" successfully recovers the overall layout of the surfing scene, but local details of the person, surfboard, and waves remain relatively blurred. With RVQ, these details become clearer, particularly in terms of object boundaries and local textures.

Removing the bottleneck branch ("Text+RVQ") results in a noticeable degradation in reconstruction performance, especially at lower SNRs. The results in Fig. 6(a) show that, at an SNR of 5 dB, removing the bottleneck branch reduces the PSNR from 26.3 dB for "Bneck+Text+RVQ" to 17.7 dB for "Text+RVQ", representing a decrease of 8.6 dB. This highlights the importance of bottleneck features in providing fundamental visual and structural information for image reconstruction, which cannot be adequately replaced by textual guidance and RVQ side information alone.

Finally, comparing "Bneck+RVQ" with "Bneck+Text+RVQ" reveals that their PSNR and SSIM values are comparable in the higher-SNR region. However, the inclusion of the textual branch helps maintain higher cross-modal semantic similarity in the lower-SNR region. As shown in Fig. 6(d), at an SNR of 2.5 dB, adding the text branch increases CLIP similarity from 0.214 for "Bneck+RVQ" to 0.235 for "Bneck+Text+RVQ". This indicates that, beyond contributing to visual reconstruction, textual information provides explicit semantic guidance that helps preserve consistency between the reconstructed image and its corresponding description.

Overall, the ablation results demonstrate that the Bneck, Text, and RVQ branches provide complementary contributions to reconstruction, particularly under low-SNR conditions. As the channel quality improves, the marginal benefits of incorporating additional information become less pronounced, although their effects may vary across evaluation metrics. A similar SNR-dependent trend is observed for layered UEP: its advantage is more evident at low SNR but diminishes as SNR increases, with equal protection even yielding slightly better results for some metrics. Fig. 6(a) illustrates how the relative performance of the two protection strategies changes with SNR. At 2.5 dB, "Bneck+Text+UEP_RVQ" achieves a PSNR of 22.8 dB, compared with 20.8 dB for "Bneck+Text+RVQ". At 10 dB, the corresponding values are 27.8 and 28.1 dB, respectively, indicating that the low-SNR advantage of unequal protection disappears at the higher SNR. These observations motivate a further investigation of channel coding rate allocation across the five information streams, namely, Bneck, Text, and the RVQ indices from the three quantization stages, as presented in the next subsection.

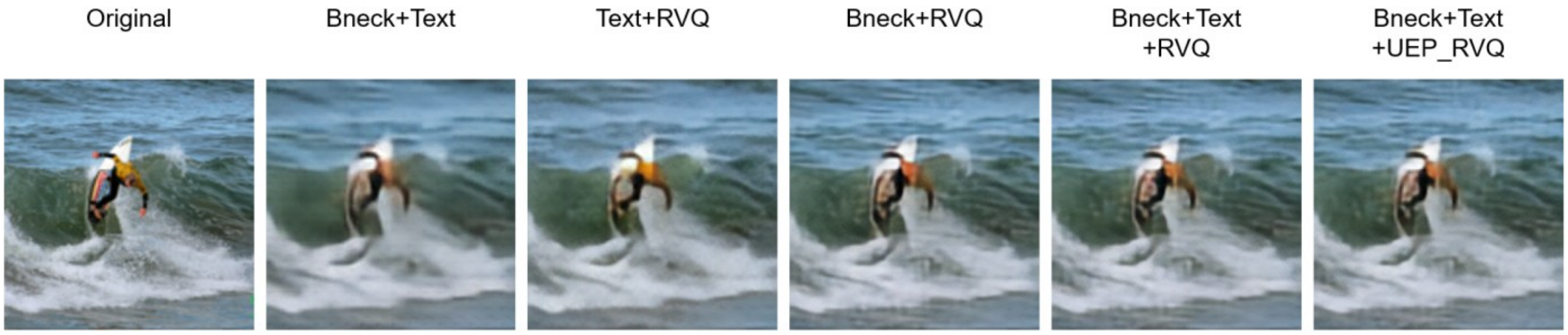

**Fig. 7.** Example image reconstructions under different conditions at SNR = 10 dB.

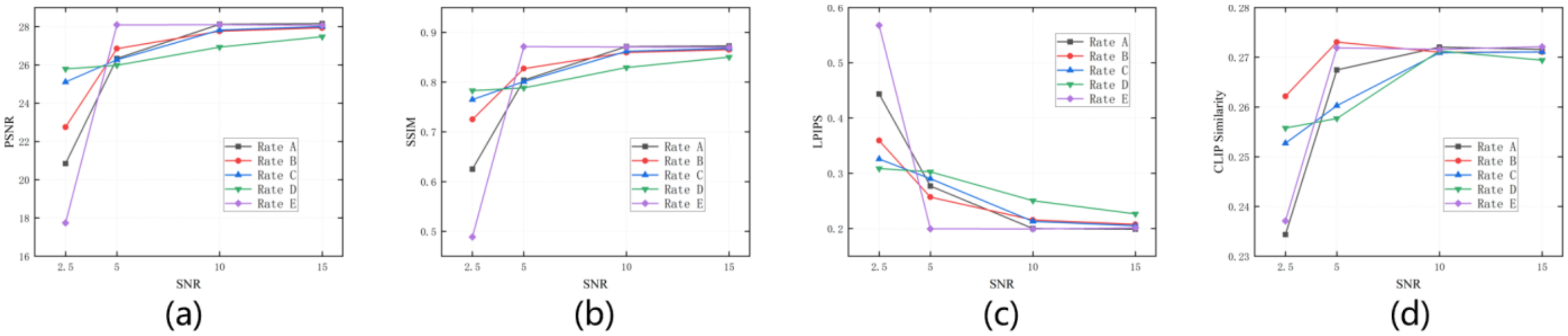

**Fig. 8.** Comparison of reconstruction performance under different code-rate allocation configurations.

### *E. Effects of Channel Coding Rate Allocation*

To further investigate the impact of unequal channel protection resource allocation on reconstruction performance, we consider five deliberately designed channel coding rate configurations, as summarized in Table II, while retaining all three information branches: “Bneck”, “Text”, and the three-stage RVQ. To ensure a fair comparison, the overall transmission budget allocated to Bneck and the three RVQ stages is kept constant across all five configurations, while the transmission overhead of the “Text” branch is excluded from this budget. The “Text” branch adopts the same channel coding rate as “Bneck” in each configuration. Accordingly, the five configurations explore different protection resource allocation strategies under a comparable transmission budget. Specifically, Configuration A applies equal protection across the three RVQ stages, whereas Configuration B introduces progressively weaker protection for later stages. Configurations C and D further vary the protection allocated to “Bneck” relative to the RVQ stages, while Configuration E assigns an identical code rate to all four budgeted streams.

Table II. Channel coding rate allocation configurations

| Config. | Bneck / Text | RVQ-1 | RVQ-2 | RVQ-3 |
|---|---|---|---|---|
| A | 0.333 | 0.5 | 0.5 | 0.5 |
| B | 0.333 | 0.333 | 0.571 | 0.8 |
| C | 0.307 | 0.5 | 0.6 | 0.75 |
| D | 0.286 | 0.667 | 0.8 | 0.8 |
| E | 0.389 | 0.389 | 0.389 | 0.389 |

The results in Fig. 8 reveal a tradeoff in channel protection resource allocation between the foundational information streams (“Bneck” and Text) and the RVQ-based visual enhancement stream. The former provides essential structural and semantic conditions for reconstruction, whereas the latter supplies complementary visual details.

In the low-SNR region, stronger protection of the "Bneck” and Text branches leads to better reconstruction performance. As the allocation shifts from Configuration B to Configuration D, more protection resources are assigned to these foundational branches. Specifically, as the channel coding rate of Bneck and Text decreases from 0.333 (config B) to 0.286(config D), the corresponding PSNR values at 2.5 dB SNR increase in ascending order: 22.8, 25.1, and 25.8, as shown in Fig. 8(a). This indicates that under severe channel conditions, maintaining the reliability of the fundamental structural and semantic information is particularly important for reconstruction.

However, this trend reverses as the channel condition improves. At 5 dB SNR, the corresponding PSNR values from Configuration B to Configuration D exhibit a descending trend, namely, 26.9, 26.3, and 25.9. This suggests that further strengthening the protection of “Bneck” and Text does not necessarily yield better reconstruction performance. Instead, allocating more protection resources to the RVQ branches becomes increasingly beneficial, as improved channel conditions allow the receiver to make more effective use of the supplementary visual information. These results demonstrate that channel protection resource allocation should account for the prevailing channel conditions. Rather than consistently favoring either the foundational information or the RVQ-based visual enhancement, an appropriate balance between their protection levels is essential for achieving robust reconstruction across different SNR regimes.

Furthermore, Configurations A and E are included to examine two uniform protection schemes: Configuration A assigns an identical code rate to all RVQ stages, whereas Configuration E applies a uniform code rate across all information streams. As shown in Fig. 8, the relative performance of these configurations varies with channel SNR, and neither consistently outperforms the other allocation strategies across all channel conditions and evaluation metrics. This further suggests that a fixed channel coding rate allocation may not be optimal under varying channel conditions. Instead, channel-state feedback could be exploited to guide the adaptive allocation of channel coding resources, motivating future research into channel-aware protection strategy learning.

## V. Conclusion

This paper proposes a generative image communication framework based on cross-modal semantic successive refinement for underwater acoustic channels. At the receiver, the received text description and bottleneck features are first fed into the Inter Predictor to generate a base estimate of the intermediate feature. The received RVQ indices are then decoded to recover visual side information, which is used by the residual refinement branch to further refine the base estimate. Finally, a one-step Pixel MeanFlow model reconstructs the RGB image from the refined intermediate feature conditioned on the text embedding. In addition, a CLIP-based text–image semantic consistency constraint is incorporated during training to encourage semantic consistency between the reconstructed image and its corresponding text description.

Experimental results show that removing the bottleneck branch leads to a substantial degradation in reconstruction quality, particularly under low-SNR conditions, while incorporating the text branch improves the semantic similarity between reconstructed images and their corresponding descriptions. The additional RVQ-based visual side information further improves reconstruction quality, especially at higher SNRs. Increasing the RVQ quantization depth and codebook size also generally improves reconstruction performance, although at the cost of increased transmission overhead. Furthermore, the channel protection experiments show that stronger protection of the bottleneck and text branches is more beneficial under low-SNR conditions. As the channel SNR increases, allocating relatively more protection resources to the RVQ stages becomes increasingly beneficial, rather than continuously strengthening the protection of the bottleneck and text branches. The results also show that unequal protection across RVQ stages improves reconstruction performance under low-SNR conditions, while its advantage diminishes at higher SNRs. These findings demonstrate the importance of adapting channel coding rate allocation to the channel conditions. Future work will investigate channel-aware adaptive protection strategies that dynamically adjust the protection levels of the bottleneck, text, and RVQ branches according to channel variations, aiming to further improve reconstruction performance under diverse underwater acoustic conditions.